%% file: VR-TERAHERTZ.tex
\documentclass[conference]{IEEEtran}

\usepackage{bbm}
\usepackage{amsmath}
\usepackage{acronym}  
\usepackage[dvips]{color}
\usepackage{epsf}
\usepackage{times}
\usepackage{epsfig}
\usepackage{graphicx}
\usepackage{epstopdf}
\usepackage{pstricks}
\usepackage{amsmath}
\usepackage{amssymb}
\usepackage{amsxtra}
\usepackage{here}
\usepackage{rawfonts}
\usepackage{times}
\usepackage{url}
\usepackage{cite}
\usepackage{caption}
\usepackage{subcaption}
\usepackage{algorithm}
\usepackage{algorithmic}
\usepackage{blindtext}
\usepackage{enumitem}
\usepackage{xcolor}
\usepackage{relsize}
\usepackage{lipsum}
\usepackage{graphicx}

\usepackage{mathtools}

\usepackage{graphics}
\usepackage{physics}
\usepackage{amssymb}
\usepackage{siunitx}

\include{newcommands}
\usepackage{multicol}

\usepackage[nomain,acronym,shortcuts]{glossaries}
\makeglossaries
\newcommand*{\acro}[3][]{\newacronym[#1]{#2}{#2}{#3}}
\include{acronyms}

\usepackage{amssymb}
\usepackage{subcaption}
\usepackage[utf8]{inputenc}
\usepackage[english]{babel}
 
\usepackage{amsthm}
 
\theoremstyle{definition}

\theoremstyle{corollary}

\theoremstyle{theorem}
\newtheorem{theorem}{Theorem}
\theoremstyle{lemma}
\newtheorem{lemma}{Lemma}
\newcommand\blfootnote[1]{%
  \begingroup
  \renewcommand\thefootnote{}\footnote{#1}
  \addtocounter{footnote}{-1}
  \endgroup
}

\begin{document}
\title{\vspace{-0.5cm}On the Reliability of Wireless Virtual Reality at Terahertz (THz) Frequencies\vspace{-0.4cm}}
\author{\IEEEauthorblockN{Christina Chaccour\IEEEauthorrefmark{1},            
        Ramy Amer\IEEEauthorrefmark{2},
        Bo Zhou\IEEEauthorrefmark{1},
     and Walid Saad\IEEEauthorrefmark{1}}
    \IEEEauthorblockA{\IEEEauthorrefmark{1}Wireless@ VT, Bradly Department of Electrical and Computer Engineering, Virginia Tech, Blacksburg, VA USA,}
    \IEEEauthorblockA{\IEEEauthorrefmark{2}CONNECT, Trinity College, University of Dublin, Ireland.}
    \IEEEauthorblockA{Emails:\{christinac, ecebo, walids\}@vt.edu, ramyr@tcd.ie \vspace{-11mm}}
    
}

\maketitle

\begin{abstract}  
Guaranteeing ultra reliable low latency communications (URLLC) with high data rates for \ac{VR} services is a key challenge to enable a dual \ac{VR} perception: visual and haptic. In this paper, a \ac{THz} cellular network is considered to provide high-rate \ac{VR} services, thus enabling a successful visual perception. For this network, guaranteeing URLLC with high rates requires overcoming the uncertainty stemming from the THz channel. To this end, the achievable reliability and latency of VR services over THz links are characterized.  In particular, a novel expression for the probability distribution function of the transmission delay is derived as a function of the system parameters. Subsequently, the \ac{E2E} delay distribution that takes into account both processing and transmission delay is found and a tractable expression of the reliability of the system is derived as a function of the THz network parameters such as the molecular absorption loss and noise, the transmitted power, and the distance between the \ac{VR} user and its respective \ac{SBS}. Numerical results show the effects of various system parameters such as the bandwidth and the region of non-negligible interference on the reliability of the system. In particular, the results show that \ac{THz} can deliver rates up to $16.4$ Gbps and a reliability of $99.999\%$ (with a delay threshold of $30$ ms) provided that the impact of the molecular absorption on the \ac{THz} links, which substantially limits the communication range of the \ac{SBS}, is alleviated by densifying the network accordingly.  
\end{abstract}
\begin{IEEEkeywords}
virtual reality (VR), terahertz, reliability, \acrfull{URLLC}.
\vspace{-0.35cm}
\end{IEEEkeywords}
\IEEEpeerreviewmaketitle
\section{Introduction}
\label{sec:intro}
\blfootnote{\noindent This research was supported by the National Science Foundation under Grant CNS-1836802.}\Acrfull{VR} is perhaps one of the most anticipated technologies of the coming decade \cite{saad2019vision}. However, relying on wired \ac{VR} systems significantly limits the \ac{VR} application domain. Instead, the deployment of wireless \ac{VR}, over cellular networks, can potentially unleash its true potential\cite{interconnected,saad2019vision}.
In order to integrate \ac{VR} services over wireless networks, it is imperative to ensure \acrfull{URLLC} \cite{interconnected}.
Predominantly, the \acrfull{E2E} delay for wireless \ac{VR} needs very low (at the order of tens of milliseconds) in order to maintain a satisfactory user experience.  Moreover, along with URLLC, wireless VR services will also require high data rates in order to deliver the $360^\circ$ content to their users. Guaranteeing high data rates with URLLC requires a major departure from classical URLLC services that were limited to low-rate sensors \cite{urllconly}. To overcome this challenge, one can explore the high bandwidth available at the terahertz (THz) frequency bands which can enable high-rate wireless access for VR \cite{coverage}. However, the reliability of the THz channel can be impeded by its susceptibility to blockage, molecular absorption, and communication range. Thus, it is imperative to understand whether THz frequencies can indeed provide an immersive VR experience by delivering URLLC at high rates.  
A number of recent works attempted to address the challenges of \ac{VR} communications \cite{interconnected, wirelessera, minzghe, urllcembb, edgecomputing}.
In \cite{interconnected}, the authors discuss the current and future trends of wireless \ac{VR} systems. The work in \cite{wirelessera} introduces a perception-based mixed-reality video streaming delivery system to provide the aggregate data rates needed for VR services. In \cite{minzghe}, the authors propose a VR model using multi-attribute utility theory to capture the 
Meanwhile, the recent works in \cite{urllcembb} and \cite{edgecomputing} study the problem of \ac{URLLC} for \ac{VR} networks. In \cite{urllcembb},  the issue of concurrent support of visual and haptic perceptions over wireless cellular networks is studied, while the work in \cite{edgecomputing} proposes a joint proactive computing and millimeter wave resource allocation scheme under latency and reliability constraints. However, these prior works \cite{minzghe, urllcembb, edgecomputing} only examine the average delays and data rates; thus reflecting limited information about the systems analyzed. In contrast, to guarantee URLLC, it is necessary to have a full view of the statistics of the delay in order to properly characterize the system's reliability. 
Last, but not least, we note that the use of \ac{THz} has recently attracted attention (e.g., see \cite{softwaredefined} and \cite{coverage}) as an enabler of high data rate applications. However, these prior works in \cite{softwaredefined} and \cite{coverage}  do not address issues of reliability or low-latency for wireless VR systems. Clearly, there is a lack in existing works that study the potential of \ac{THz} frequencies to deliver high-data rate \ac{VR} services while providing \ac{URLLC}.\\
\indent The main contribution of this paper is to analyse the delay and reliability performance of a cellular network operating at \ac{THz} frequencies and servicing \ac{VR} users. The ultimate goal is to assess how and when a \ac{THz} network can meet the dual requirements of the \ac{QoS} of a \ac{VR} user, in terms of \ac{URLLC} and high data rates. In particular, we introduce a novel \ac{VR} model based on a \ac{MHCPP}. In the studied model, each VR user sends a request to its respective \acrfull{SBS}, which induces an \ac{E2E} delay that includes the delay needed to process the VR images and the transmission delay over the THz links. Based on this model, to find the \ac{CDF} of the \ac{E2E} delay, we derive the \ac{PDF} of the transmission delay in a dense \ac{THz} network.  Subsequently, this allows us to derive the reliability of the system and to characterize the network parameters that affect this reliability. To our best knowledge, \textit{this is the first work that analyzes the reliability and latency achieved by VR services over a \ac{THz} cellular network.} Simulation results show that reliability is mainly affected by the transmission delay, which can be significantly reduced by providing the system with higher bandwidth or through densifying the network to maintain short-range communication between SBSs and users, thus overcoming the molecular absorption effects and guaranteeing reliability.
\section{System Model}
Consider the downlink of a small cell network servicing a set $\mathcal{V}$ of $V$ wireless VR users via a set of \acp{SBS} distributed in a confined indoor area according to an isotropic homogeneous \ac{MHCPP}  with intensity $\eta$ and a minimum distance $r$ \cite{haenggi2012stochastic}. This process is a special thinning of the \ac{PPP} in which the nodes are forbidden to be closer than a minimum distance $r$. 
Here, the parameter $r$ indicates that the distance between adjacent nodes cannot be arbitrarily small in the real-life phenomenon. Hence, this process can adequately capture the distribution of \ac{VR} \acp{SBS} in a confined area. The SBSs can also perform \ac{MEC} functions for VR purposes.
\vspace{-0.2cm} 
\subsection{Wireless Capacity}
\vspace{-0.1cm} 
We consider an arbitrary \ac{VR} user in $\mathcal{V}$ that is at a constant distance $d_0$ from its respective \ac{SBS}. Hence, the chosen \ac{VR} user and its respective \ac{SBS} are referred to as \textit{tagged} receiver and transmitter respectively. The interference surrounding this VR user stems from a set $\mathcal{M}$ of $M$ non-negligible interfering \acp{SBS} that are located within a radius of $\Omega$ around this user. This interference occurs because we consider a highly dense THz network whose SBSs are located at very close proximity. Henceforth, the SBSs that are at a distance $d$ $\geq \Omega$ add no interference on the link between the VR user to its associated SBS. As shown in \cite{7390991}, the signal propagation at the THz-band is mainly affected by molecular absorption, which results in molecular absorption loss and molecular absorption noise. Given that the distance between the \ac{VR} user and its respective \ac{SBS} is short in our model, we consider a \ac{LoS} link only, as also done in \cite{softwaredefined}. 
Consequently, the total path loss affecting the transmitted signal between the SBS and the \ac{VR} user will be given by \cite{7390991}:
\vspace{-0.4cm}
\begin{equation}
L(f,d)=L_s(f,d)L_m(f,d)=\left(\frac{4\pi f d}{c}\right)^2 \frac{1}{\tau(f,d)},
\end{equation}
where $L_s(f,d)=(\frac{4\pi f d}{c})^2$ is the free-space propagation loss, $L_m(f,d)=\frac{1}{\tau(f,d)}$ is the molecular absorption loss, $f$ is the operating frequency, $d$ is the distance between the VR user and the SBS,  $c$ is the speed of light, and $\tau(f,d)$ is the transmittance of the medium following the Beer-Lambert law, i.e., $\tau(f,d)\approx {\rm exp}(-K(f)d)$, where $K(f)$ is the overall absorption coefficient of the medium.
 Let $\boldsymbol{d}\triangleq(d_i)_{i=0,1,\cdots,M}$ be a row vector, where $d_0$ denotes the distance between the VR user and the associated SBS, and $d_i$ denotes the distance between the VR user and the interfering SBS $i\in\mathcal{M}$. Let $\boldsymbol{p}\triangleq(p_i)_{i=0,1,\cdots,M}$ be a row vector, where $p_{0}$ denotes the transmission power of the SBS servicing the considered VR user, and $p_i$ denotes the transmission power of the interference from any other SBS $i\in\mathcal{M}$. The total noise power is the sum of the molecular absorption noise and the Johnson-Nyquist noise generated by thermal agitation of electrons in conductors.
 Consequently, the total noise power at the receiver can be given by \cite{7390991}:
 \vspace{-3mm}
\begin{equation}
N(\boldsymbol{d}, p_i,f)=N_0+\sum_{i=1}^{M}p_iA_od_i^{-2}(1-e^{-K(f)d_i}),
\vspace{-0.2cm}
\end{equation} 
where $N_0=k_B T +p_{T_0}A_0d_0^{-2} (1-e^{-{K(f)d_0}})$, $k_B$ is Botlzmann constant, $T$ is the temperature in Kelvin, and $A_0=\frac{c^2}{16 {\pi}^2 f^2}$.
Furthermore, accounting for the total path loss affecting the transmitted signal, the aggregate interference will be: $I(\boldsymbol{d}, {p_{i}},f)=\sum_{i=1}^{M}p_iA_od_i^{-2}e^{-K(f)d_i}$.
The instantaneous frequency-dependent \ac{SINR} is then given by:
\begin{equation}
S(\boldsymbol{d}, \boldsymbol{p},f)=\frac{p^{\textrm{RX}}_0(d_0,p_{0}, f)}{I(\boldsymbol{d}, p_i,f)+N(\boldsymbol{d}, p_i,f)},
\end{equation}
where $p^{\textrm{RX}}_0$ is the received power at the VR user from its associated SBS. Substituting each of the received power, noise and interference term results in the following SINR:
\begin{equation}
S(\boldsymbol{d}, \boldsymbol{p},f)=\frac{p_{0}A_0d_0^{-2}e^{-K(f)d_0}}{N_0+\sum_{i=1}^{M}p_iA_od_i^{-2}}.
\end{equation}
Hence, the capacity of the channel can be written as:
\begin{equation}
\label{capacity}
C(\boldsymbol{d}, \boldsymbol{p},f)=W {\log}_2\left(1+\frac{p_{0}A_0d_0^{-2}e^{-K(f)d_0}}{N_0+\sum_{i=1}^{M}p_iA_od_i^{-2}}\right),
\end{equation}
where $W$ is the bandwidth.
\subsection{Interference Analysis}
From \eqref{capacity}, we can see that the only random factor is the second term in the denominator which corresponds to the interfering signals.
For technical tractability, following \cite{7390991}, we assume that this term tends to a normal distribution \cite{pechinkin1973convergence}. Note that, it has been shown in \cite{7390991} that such an approximation is realistic.
Furthermore, finding the mean and variance of this term will allow us to characterise the \ac{PDF} of this random interference signal, as follows:
\begin{equation}
\label{eq:interference}
g(I) = \frac{1}{\sqrt{2\mathcal{\pi}}\sigma_I}{\rm exp}{\left(-\frac{(I-\mu_I)^2}{2\sigma_I^2}\right)},
\end{equation}
where  $\mu_I$ and $\sigma_I^2$ are the mean and variance of the interference, respectively, and are given by \cite{7390991}:
\begin{align}
\mu_I&=pA_0{\left(\frac{\ln(\Omega)-\ln(r)}{\Omega^2-r^2}\right)}{ \left(\frac{\pi \Omega^2\eta}{2}\right)},\\
\sigma_I^2&=(pA_0)^2{\left(\frac{\pi \Omega^2 \eta}{2}\right)}{\left(\frac{1}{2r^2 \Omega^2}\right)},
\end{align}
where $r$ is the minimum distance of the \ac{MHCPP}, $\Omega$ is the region of non-negligible interference, and the subscript $i$ in $p_i$ is omitted given that the \acp{SBS} are assumed to have the same  transmission power.
As shown  in \cite{7390991}, $\mu_I$ and $\sigma_I$ can be derived based upon the Poisson approximation for the distances between the tagged receiver and the interferers. 
Given the high bandwidth available at the \ac{THz} band can provide high-rate wireless \ac{VR}, however, it is necessary to analyze whether this network can provide \ac{URLLC}. Next, we will analyze the delay of the considered system and, then, leverage this analysis to define reliability. \vspace{-0.2cm}
\section{Reliability Analysis}
\subsection{Delay Analysis}
\begin{figure}[!t]
\vspace{-0.4cm}
\centering
    \includegraphics[width=0.4\textwidth]{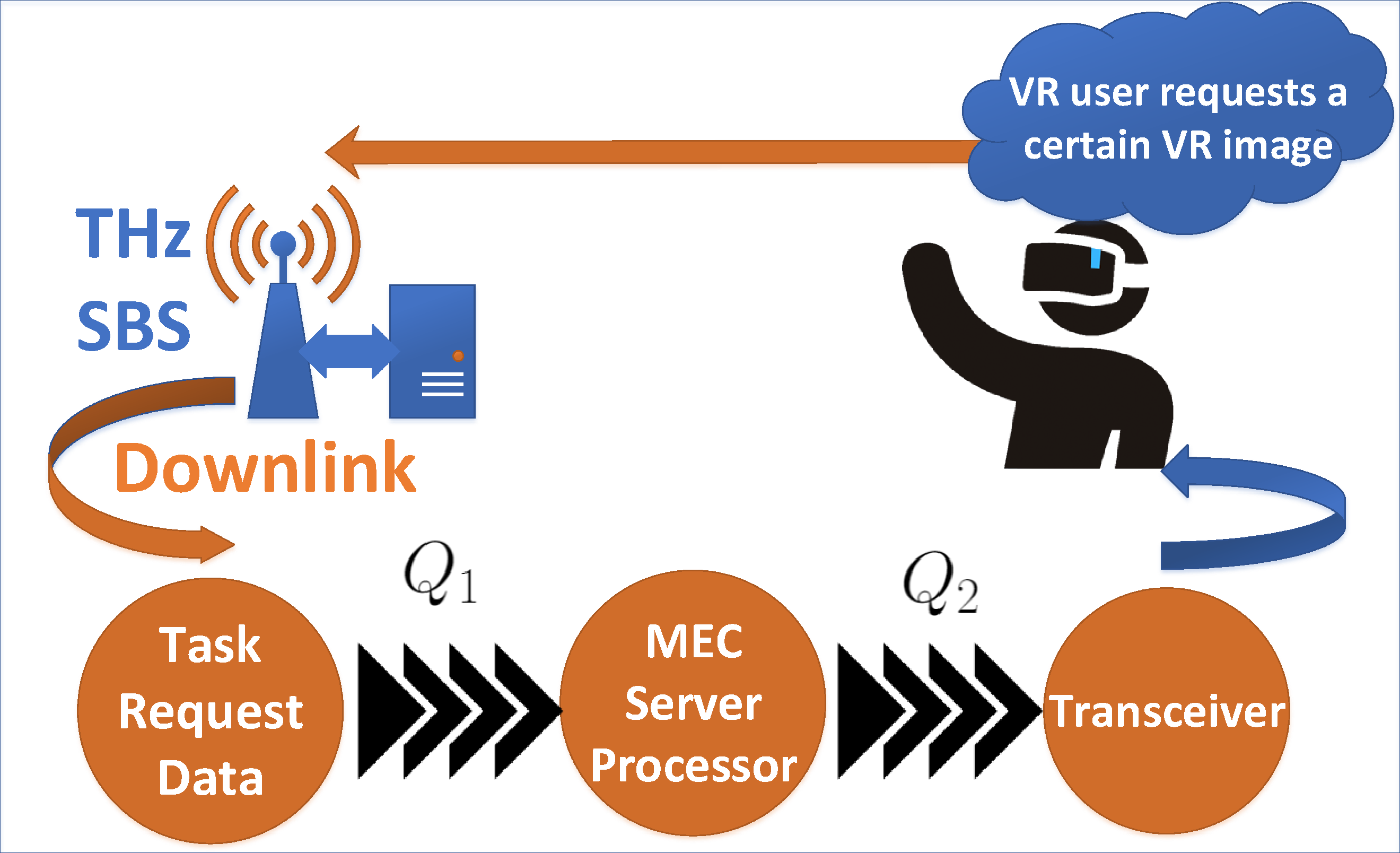}
    \caption{\small{Illustrative example of our system model.}}
    \label{fig:boat1}
\vspace{-0.4 cm}
\end{figure} 
The service model of the VR image request in our wireless VR system is illustrated in Fig.~\ref{fig:boat1}. As shown, once a  VR user requests a VR image, this request will go through two queues: a first queue, $Q_1$, that pertains to processing a $360^{\circ}$ VR image, and a second queue, $Q_2$, that pertains to storing and transmitting the VR images over the wireless THz channel. Here, we assume that the time for sending a request by the VR user is negligible.
Hence, for each \ac{VR} image request, the total delay depends on the waiting and the processing time at $Q_1$ and the waiting time and VR transmission delay at $Q_2$.
\indent We assume that a \ac{VR} image request follows a Poisson arrival process with mean rate $\lambda_1$. The buffer of the processor is assumed to be of infinite size and the MEC processor at the SBS adopts the first-come, first-serve (FCFS) policy. 
 The service time for each request follows an exponential distribution with rate parameter $\mu_1 > \lambda_1$, to guarantee the stability of the first queue $Q_1$. Thus, we can see that the queue $Q_1$ is an M/M/1 queue. 
According to Burke's theorem \cite{queueingtheory}, when the service rate is larger than the arrival rate for an M/M/1 queue, then the departure process at steady state is a Poisson process with the same arrival rate.
Hence, the arrival of requests to $Q_2$ also follows a Poisson process with rate $\lambda_2=\mu_1$. 
Similar to $Q_1$, we assume an infinite buffer size and an FCFS policy for $Q_2$. 
Note that, the service time of $Q_2$ is the transmission time of the SBS, that depends on the random wireless THz channels. 
Thus, different from $Q_1$, the second queue $Q_2$ is an M/G/1 queue.
Our goal is to study when and how the proposed \ac{THz} system can guarantee the dual \ac{QoS} requirement required by \ac{VR}, i.e.  visual and haptic perceptions. This dual perception requires a high data rate link for visual perception and a low latency communication for the haptic. 
Under favorable channel conditions, \ac{THz} is capable of providing high rate links, however, providing URLLC may be challenging. Hence, our key step is to define the reliability of this system and study the performance of the \ac{VR} network. This performance analysis will shed light on the capability of \ac{THz} to provide a dual-metric performance for VR.
To analyze reliability, next, we first derive the \ac{PDF} of the transmission delay of our \ac{THz} network. This expression will then be used to derive the \ac{CDF} of the \ac{E2E} delay characterizing the reliability of the system and the \ac{QoS}.
It is important to note that reliability cannot be defined merely on average values of delays as done in \cite{minzghe} and \cite{interconnected}. Given the stringent requirements of \ac{VR} services, a full view on the statistics of the delay must be to be taken into account in order to design a system capable of withstanding extreme and infrequently occurring events such as a sudden user movement that changes its distance from its respective \ac{SBS} or a sudden blockage between the user and the \ac{SBS} which can impact reliability.
\vspace{-0.1cm}
\subsection{Reliability Analysis}
The reliability of the considered wireless \ac{VR} system can be defined as a guarantee that the \ac{E2E} delay can be maintained below a target threshold $\delta$. Formally, reliability is the probability that the \ac{E2E} delay -- defined as the delay incurred between the time the \ac{VR} user requests a VR image to the time the image is received -- remains below $\delta$. Hence, the system is guaranteed to have high reliability when this probability is high and tends to 1.
For our model in Fig.~\ref{fig:boat1}, given that $Q_1$ is an M/M/1 queue, the \ac{PDF} of the total waiting time at $Q_1$ will be \cite{queueingtheory}:
\begin{equation}
\label{MM1}
\psi_1(t)=(\mu_1-\lambda_1)\exp\left(-(\mu_1-\lambda_1)t\right).
\end{equation}
Moreover, given that $Q_2$ is an M/G/1 queue and that the queuing and service time of an M/G/1 queue are independent, we can derive that the \ac{CDF} of the total waiting time:
\begin{equation}
\label{Q2conv}
\Psi_2(t)=\Psi_{Q_2}(t)*\psi_T(t),
\end{equation}
where $*$ is the convolution operator, $\Psi_{Q_2}(t)$ is the \ac{CDF} of the queuing time at $Q_2$ and $\psi_T(t)$ is the \ac{PDF} of the transmission delay. The \ac{CDF} of the total queuing time at $Q_2$ will be \cite{queueingtheory}\vspace{-0.3cm}:
\begin{equation}
\Psi_{Q_2(t)}=(1-\rho)\sum_{n=0}^{\Gamma}{\left[\rho^{n}R^{(n)}(t)\right]},
\vspace{-0.3cm}
\end{equation}
where $\rho=\frac{\lambda_2}{\mu_2}$ is the utilization factor, $\lambda_2$ and $\mu_2$ are the arrival and average transmission rates of $Q_2$, respectively. Here, $\Gamma$ is the number of states that the queue has went through, i.e., the number of packets that has passed through the queue during a certain amount of time and $R^{(n)}(t)$ is the \ac{CDF} of the residual service time after the $n$-th state. Note that $R^{(n)}(t)$ can be computed by obtaining the residual service time distribution $R(t)$ after $n$ packets, $R(t)=\int_{0}^{t} \mu_2 (1-\psi_T(x))dx$, where $t$ is the time of an arbitrary arrival, given that the arrival occurs when the server is busy.
To evaluate the \ac{PDF} of the \ac{E2E} delay, we need the PDF of the transmission delay which is found next:
\begin{lemma}
The \ac{PDF} of the transmission delay is given by: \begin{equation}\label{PDFTX}
\psi_T(\alpha)=\frac{\zeta}{\sqrt{2\pi}\sigma_I}\exp(\frac{(\Upsilon -\mu_I)^2}{2\sigma_I^2}),
\vspace{-3mm}
\end{equation}
where 
\begin{eqnarray}
&&\zeta=\frac{\ln{(2)}(p^{\textrm{RX}}_0L)2^{\frac{L}{W\alpha}}}{W\mathcal{\alpha}^2 (2^{\frac{L}{W\alpha}}-1)^2},\\
&&\Upsilon=\frac{(1-2^{\frac{L}{W\alpha}})N_0+p^{\textrm{RX}}_0}{2^{\frac{L}{W\alpha}}-1}, 
\end{eqnarray} 
where $L$ is the packet size and $\alpha$ is the transmission delay.
\end{lemma}
\begin{IEEEproof}
Based on \eqref{capacity}, we express the capacity in terms of $L$ and $\alpha$ as:
\begin{align}\label{eqn:proofalpha}
&C=\frac{L}{\alpha}=W {\log}_2\left(1+\frac{p_{0}A_0d_0^{-2}e^{-K(f)d_0}}{N_0+\sum_{i=1}^{M}pA_od_i^{-2}}\right),
\end{align}
We can see that the only random term in \eqref{eqn:proofalpha} is the interference that is assumed to follow a normal distribution. Subsequently, we can express the interference in terms of the transmission delay $\alpha$ as follows:
\begin{align}
&\sum_{i=1}^{M}pA_od_i^{-2}=\frac{N_0\left(1-2^\frac{L}{W\alpha}\right)+p^{\textrm{RX}}_0}{2^\frac{L}{W\alpha}-1}.
\vspace{-0.5cm}
\end{align}
By applying the transform for \ac{PDF}s $g(y)=g(x)\frac{\partial x}{\partial y}$ we can find the \ac{PDF} of transmission delay by transforming the \ac{PDF} of interference accordingly.
We let $\Upsilon$ represent the interference and $\zeta$ its derivative with respect to the transmission delay. Then, we have:
 \vspace{-0.22cm}
\begin{align}
\Upsilon&=\sum_{i=1}^{M}p_iA_od_i^{-2},\\
\zeta&=\frac{d\Upsilon}{d\alpha}=\frac{\ln\left(2\right)\ln{\cdot}2^\frac{L}{W\alpha}}{W\alpha^2\left(2^\frac{L}{W\alpha}-1\right)} \nonumber\\
&\quad\quad\quad\quad +\dfrac{\ln\left(2\right)L\left(N_0\left(1-2^\frac{L}{W\alpha}\right)+p^{\textrm{RX}}_0\right){\cdot}2^\frac{L}{W\alpha}}{W\alpha^2\left(2^\frac{L}{W\alpha}-1\right)^2}\nonumber\\
&\quad\quad\quad=\dfrac{\ln\left(2\right)L p^{\textrm{RX}}_0{\cdot}2^\frac{L}{W\alpha}}{W\alpha^2\left(2^\frac{L}{W\alpha}-1\right)^2}.
\end{align}
\vspace{-0.15cm}
Hence, the transmission delay \ac{PDF} will be:
\begin{align}
\label{eq:transmission}
\psi_T(\alpha)&= g(\Upsilon)\frac{d\Upsilon}{d\alpha}=\zeta g(\Upsilon)\nonumber\\
&=\frac{\zeta}{\sqrt{2\pi}\sigma_I}\exp(\frac{(\Upsilon -\mu_I)^2}{2\sigma_I^2}).
\end{align}
\end{IEEEproof} 
It is important to note that the \ac{PDF} in \eqref{eq:transmission} does not follow a normal distribution since both $\Upsilon$ and $\zeta$ depend on the transmission delay $\alpha$. 
Burke's Theorem allows us to infer that $Q_1$ and $Q_2$ are independent; therefore, the \ac{CDF} of the \ac{E2E} delay can be expressed as the convolution of the \ac{PDF} of the total waiting time in $Q_1$ and the \ac{CDF} of the total waiting time in  $Q_2$. By using the dynamics of \eqref{MM1} and \eqref{Q2conv}, the CDF of the \ac{E2E} delay can formally expressed in the following theorem which is a direct result of Lemma 1.
\begin{theorem}
The \ac{CDF} of the \ac{E2E} delay $T_e$ is given by:
\begin{align}
\label{eq:Theorem1}
\Phi(t)= P(T_e\leq t) &  =\psi_1(t)*\Psi_2(t) \nonumber\\
&=\psi_1(t)*\left(\Psi_{Q_2}(t)*\psi_T(t)\right)\nonumber\\
 &=(\mu_1-\lambda_1)\exp(-(\mu_1-\lambda_1)t)\nonumber\\
 &*\left((1-\rho)\sum_{n=0}^{\Gamma}(\rho^{\Gamma}R^{(n)}(t))\right)\nonumber\\
     &*\left(\frac{\zeta}{\sqrt{2\pi}\sigma_I}\exp(\frac{(\Upsilon -\mu_I)^2}{2\sigma_I^2}) \right). 
\end{align}
\end{theorem}
\noindent Consequently, the \emph{reliability} can be defined as the probability of the \ac{E2E} delay not exceeding a certain threshold $\delta$, i.e.,
\begin{equation}
\label{eq:rel}
\varrho=P(T_e\leq\delta)=\Phi(\delta).
\end{equation}
The reliability in \eqref{eq:rel} allows a tractable characterization of the reliability of the VR system shown in Fig. 1, as function of the THz channel parameters. Furthermore, from Theorem 1, we can first see that the queuing time of $Q_2$ depends on the residual service time \ac{CDF} and hence on the transmission delay. Also, given that the processing speed of the \ac{MEC} servers can be considerably high, the \ac{E2E} delay will often be dominated by the transmission delay of \ac{THz}. Moreover, in general, all the key parameters that have a high impact on the transmission delay will have a higher impact on reliability. One of the most important key parameters is the distance $d_0$ between the \ac{VR} user and its respective SBS; this follows from the fact that the molecular absorption loss gets significantly higher when the distance increases, which limits the communication range of THz SBSs to very few meters. Indeed, the \ac{THz} reliability will deteriorate drastically if the distance between the \ac{VR} user and its respective \ac{SBS} increased. 
Given that the \ac{QoS} of a VR application is a function of the reliability, i.e., it is the reliability of the system throughout the worst case scenario. VR users' immersion and experience will depend significantly on the reliability. Therefore, maintaining reliability is a necessary condition to guarantee the \ac{QoS} for the user, thus increasing its satisfaction and yielding it a seamless experience.
\section{Numerical Results} 
\begin{figure}[!t]
\vspace{-0.4cm}
\begin{centering}
    \includegraphics[width=0.350\textwidth]{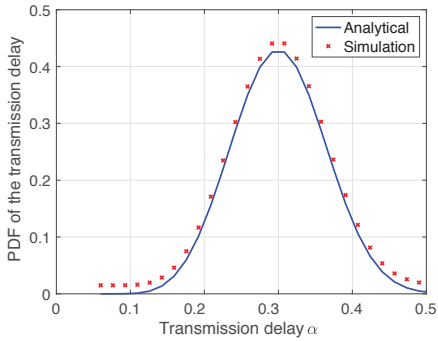}
    \caption{\small{PDF of the transmission delay.}}
    \label{fig:pdf_tx_time}
\end{centering}
\vspace{-0.4 cm}
\end{figure}

\begin{figure}[h]
	\begin{minipage}{0.49\linewidth}
		\centering
		\includegraphics[scale=0.3]{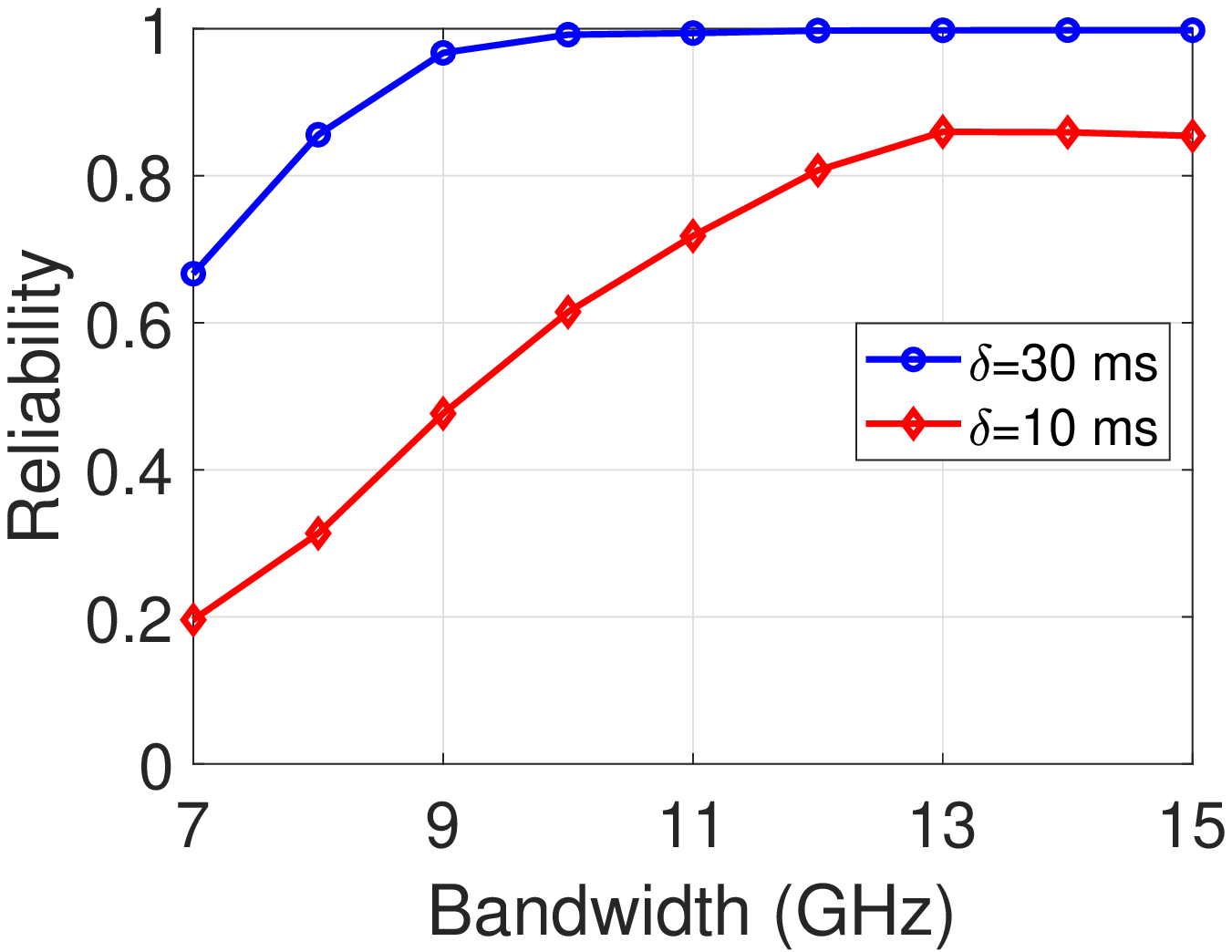}
		\subcaption{}    \label{fig:RelBandwidth}
	\end{minipage}
	\begin{minipage}{0.49\linewidth}
		\centering
		\includegraphics[scale=0.3]{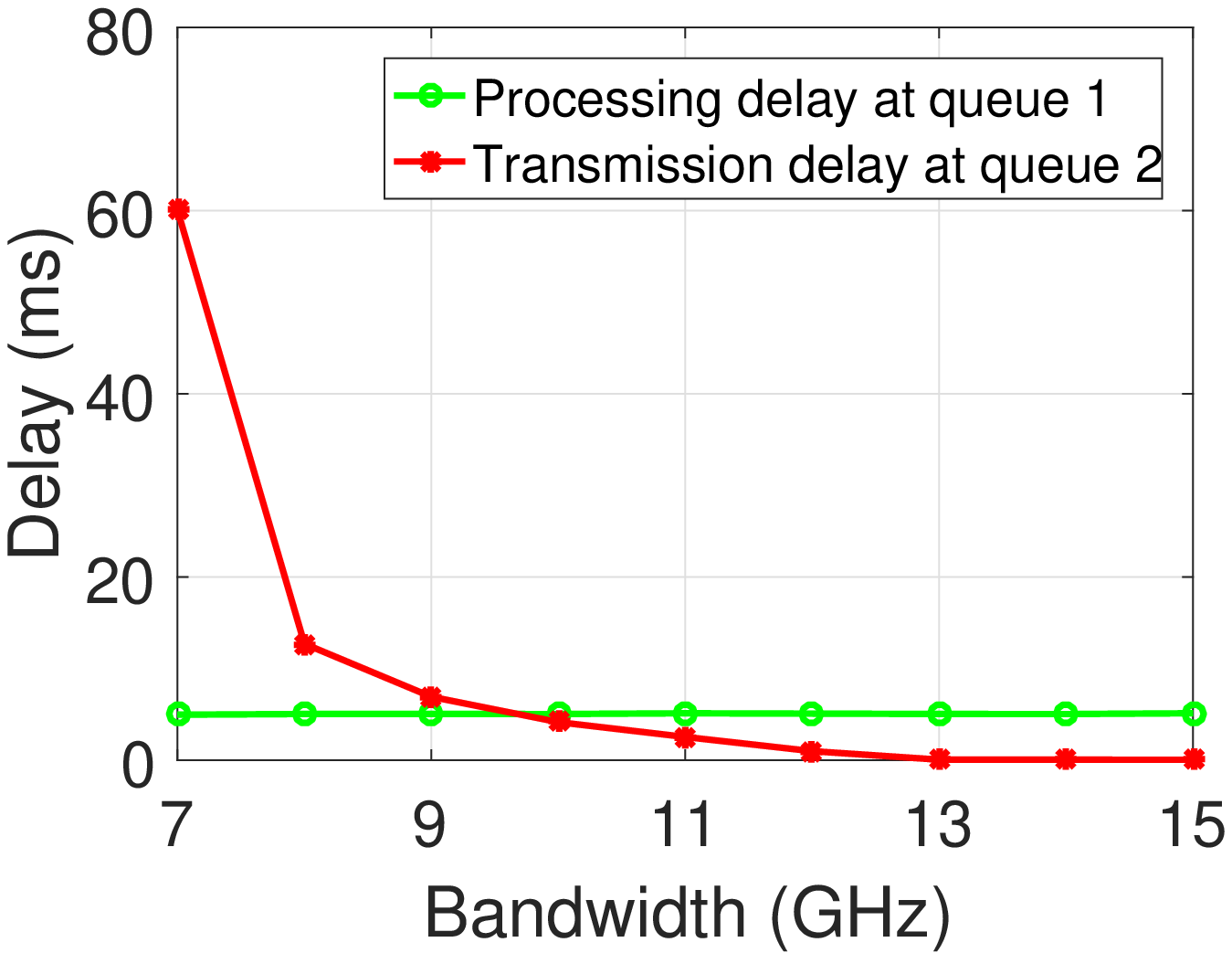}
		\subcaption{}    \label{fig:DelBandwidth}
	\end{minipage}
	\caption{\small{Effect of bandwidth on the achievable performance (a) Reliability versus bandwidth, (b) Delay versus bandwidth.}}  \label{fig:performance}
	\vspace{-0.3cm}
\end{figure}

\begin{figure} [t!]
\vspace{-0.3cm}
\begin{centering}
    \includegraphics[width=0.35\textwidth]{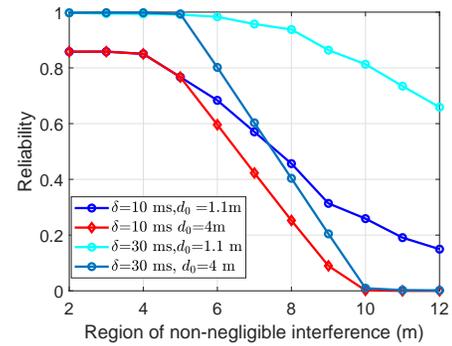}
    \caption{\small{Reliability versus region of non-negligible interference.}}
    \label{fig:region}
\end{centering}
\vspace{-0.3cm}
\end{figure}

For our simulations, we consider the following parameters: $T= \SI{300}{K}$, $p=\SI{1}{W}$, $L= \SI{10}{Mbits}$, $f=\SI{1}{THz}$, $K(f)= \SI{0.0016}{m^{-1}}$ with $1\%$ of water vapor molecules as in \cite{absorption}, $\lambda_1=\SI{0.1}{packets/s}$, and $\mu_1=\SI{2}{Gbps}$. These values are chosen to comply with existing VR processing units such as the GEFORCE RTX 2080 Ti \cite{geforce}. The SBSs are deployed in an indoor area modeled as a square of size $\SI{20}{m}\times\SI{20}{m}$. All statistical results are averaged over a large number of independent runs.\\ 
\indent Fig.~\ref{fig:pdf_tx_time} shows that the simulation results match the distribution of the analytical result derived in \eqref{PDFTX}. The small gap between the analytical and simulation results stems from the normal distribution of the interference.\\
\indent Fig.~\ref{fig:performance} (a) shows the prominent effect of the bandwidth on the reliability at different $\delta$. We can see that the reliability monotonically increases with both the bandwidth and reliability threshold $\delta$. Subsequently, we can see that, in order to achieve a reliability of $99.999\%$ at $\delta$ =\SI{30}{ms}, we need a bandwidth of $\SI{10}{GHz}$. For our THz system, this corresponds to a data rate of $\SI{16.4}{Gbps}$. Clearly, the target reliability for VR services can be achieved with high rates at THz frequencies, assuming sufficient bandwidth is available.
Furthermore, the reliability for $\delta=\SI{10}{ms}$ saturates around $W=\SI{13}{GHz}$. This is due to the fact that the delay in $Q_2$ has reached a point where it is equal to the delay in $Q_1$ in Fig.\ref{fig:performance}(b); after that point, the delay becomes dominated by the delay in $Q_1$ rather than $Q_2$. Furthermore, a very high reliability cannot be achieved for $\delta=\SI{10}{ms}$ in Fig.~\ref{fig:performance} (a) due to the delay at $Q_1$. Hence, the limitation in reliability for a very low threshold, namely $\delta=\SI{10}{ms}$ is mainly a result of the processing speed at the MEC server. Also, it is important to note that before the point of saturation, the reliability is mainly dominated by the transmission delay which confirms the result obtained in \eqref{eq:Theorem1}.\\
\indent In Fig.~\ref{fig:region}, we show how the reliability varies as a function of the region of non-negligible interference. We can see that, when the distance between the \ac{VR} user and the \ac{SBS} increases, the region of non-negligible interference $\Omega$ has a higher impact on the reliability, and the drop of reliability is sharper. This phenomenon is observed regardless of the reliability threshold $\delta$. Hence, even though the user can achieve high reliability, the dependence of the molecular absorption on distance limits the user to a very short distance to its respective \ac{SBS}. Thus, the \ac{VR} user can guarantee reliability regardless of the interference surrounding it, given that it is at a proximity from the respective \ac{SBS}.
\section{Conclusion}
In this paper, we have studied the reliability of \ac{VR} services deployed in a \ac{THz} cellular network. To obtain an expression for the end-to-end delay and reliability, we have proposed a model based on a two tandem queue. We have derived the \ac{PDF} of the transmission delay of a \ac{THz} cellular network, based on which, we have derived the \ac{E2E} delay expression along with the reliability of this system.  We have shown that operating at THz frequencies can potentially enable \ac{VR} services to have high reliability and high rates when provided with high bandwidth and proximity to the respective \ac{SBS}. Hence, the design of these networks requires managing a tradeoff between the bandwidth used and the average proximity of the user to the respective \ac{SBS}.
\bibliographystyle{IEEEtran}
\bibliography{bibliography}
\end{document}

%% file: acronyms.tex
\acro{D2D}{device-to-device}
\acro{SIR}{signal-to-interference-ratio}
\acro{SINR}{signal-to-interference-plus-noise-ratio}
\acro{PCP}{Poisson cluster process}
\acro{CoMP}{coordinated multi-point}
\acro{BS}{base station} 
\acro{MD-CoMP}{macrodiversity CoMP transmission}
\acro{MAC}{medium-access-control}
\acro{JT-CoMP}{joint transmission CoMP}
\acro{CoMP-JT}{coordinated multipoint joint transmission}
\acro{SBS}{small base station}
\acro{MDSD}{multiple devices to a single device}
\acro{MDS}{maximum distance separable}
\acro{SCN}{small cell network}
\acro{PPP}{Poisson point process}
\acro{TCP}{Thomas cluster process}
\acro{CSI}{channel state information}
\acro{PDF}{probability distribution function}
\acro{PMF}{probability mass function}
\acro{RV}{random variable}
\acro{i.i.d.}{independently and identically distributed}
\acro{MBMS}{multimedia broadcasting multicasting service}
\acro{EE}{energy efficiency}
\acro{HCP}{hard-core placement}
\acro{CCDF}{complementary cumulative distribution function}
\acro{CDF}{cumulative distribution function}
\acro{PC}{probabilistic caching}
\acro{RC}{random caching}
\acro{CPF}{caching popular files} 
\acro{PGFL}{probability generating functional}
\acro{KKT}{Karush-Kuhn-Tucker}
\acro{PGF}{point generating function}
\acro{SCA}{successive convex approximation}
\acro{HD}{high-definition}
\acro{FHD}{full-high-definition}
\acro{UHD}{ultra-high-definition}
\acro{VR}{virtual reality}
\acro{AR}{augmented reality}
\acro{5G}{fifth-generation}
\acro{QoS}{quality-of-service}
\acro{QoE}{quality-of-experience}
\acro{IoT}{internet of things}
\acro{MHCPP}{Matern hardcore point process}
\acro{LoS}{line-of-sight}
\acro{NLoS}{non-line-of-sight}
\acro{PSD}{power spectral density}
\acro{MEC}{mobile edge computing}
\acro{E2E}{end-to-end}
\acro{THz}{terahertz}
\acro{CLT}{central limit theorem}
\acro{HQ}{High Quality}
\acro{eMBB}{enhanced mobile broadband}
\acro{URLLC}{ultra reliable low latency communications}
\acro{mmWave}{millimeter wave}